\documentstyle[aps,prb,epsfig,multicol]{revtex}
\begin{document}
\draft
\title{Prominent bulk pinning effect in the MgB$_2$ superconductor}
\author{Mun-Seog Kim, C. U. Jung, Min-Seok Park, S. Y. Lee,
        Kijoon H. P. Kim, W. N. Kang, and Sung-Ik Lee}
\address{National Creative Research Initiative Center for Superconductivity
         and Department of Physics, Pohang University of Science and Technology,
         Pohang 790-784, Republic of Korea}
\date{\today}
\maketitle
\begin{abstract}
We report the magnetic-field dependence of the irreversible magnetization of the recently
discovered binary superconductor MgB$_{2}$.  For the temperature region of
$T< 0.9T_c$, the contribution of the bulk pinning to the magnetization overwhelms
that of the surface pinning. This was evident from the fact that
the magnetization curves, $M(H)$, were well described by the critical-state model
without considering the reversible magnetization and the surface pinning effect. 
It was also found that the $M(H)$
curves at various temperatures scaled when the field and the magnetization were
normalized by the characteristic scaling factors $H^\ast(T)$ and $M^\ast(T)$,
respectively. This feature suggests that the pinning mechanism determining the hysteresis
in $M(H)$ is unique below $T=T_c$.
\end{abstract}

\begin{multicols}{2}
\section{introduction}
In the mixed state, the magnetization of superconductors is a combination of two
different contributions, $M_{\rm eq}$ and $M_{\rm irr}$. $M_{\rm eq}$ is the
equilibrium (or reversible) magnetization\cite{hao1}
and $M_{\rm irr}$ is the irreversible magnetization.
The former is caused by the equilibrium
surface current. The latter arises from the surface (Bean-Livingston)
barrier effect,\cite{bean2} as well as the bulk pinning due to
the interaction between vortices and various defects within the
superconductor.
The surface barrier originates from the competition between two forces,
(a) an attractive interaction between a vortex and its image vortex and
(b) a repulsive  interaction between a vortex and the surface shielding current.
For high-$T_c$ cuprate superconductors, the irreversible
magnetization at low temperatures is dominated by the bulk pinning.
However, the role of the surface barrier effect becomes significant as the
temperature increases.\cite{dewhurst7}

Recently, superconductivity in a non-cuprate binary compound MgB$_2$ was
discovered by Akimitsu {\em et al.}\cite{akimitsu1} This  material
is known to be type II superconductor with the Ginzburg-Landau parameter 
$\kappa \sim$ 26 and $T_c \simeq$ 40 K,\cite{finnemore7} and various experimental 
studies\cite{budko7,wnkang7,larbalestier7,gabino7,slusky7,lorenz7,canfield7,takano7}
have been carried out to elucidate the fundamental properties of this new superconductor.

In the mixed state, the magnetic behavior of MgB$_2$ has been known to resemble
that of the conventional superconductors such as Nb-Ti and NbSn$_3$.
Larbalestier {\em et al}.\cite{larbalestier7} showed that the parameter 
$H^{0.25}\Delta M^{0.5}$ is linear in $H$ over a wide range of temperature 
as in Nb$_3$Sn, where the $\Delta M(\propto J_c)$ is the magnetic hysteresis in $M(H)$. 
They also reported the proportionality of the irreversible field $H_{\rm irr}(T)$ 
and the upper-critical field $H_{c2}(T)$, 
which are usually independent in high-$T_c$ cuprates.

In this work, we measured magnetization $M(H)$ of MgB$_2$ superconductor
as a function of the external magnetic field to elucidate its pinning properties
in detail. We found that the $M(H)$ curves for various temperatures
can be describe by the exponential critical state model.\cite{feitz7}
From this analysis, we present evidence of the significant role of bulk pinning 
in this system even up to $T/T_c\sim 0.9$, which is contrary to the case of 
high-$T_c$ cuprates.

\section{Experimental}
A commercially available powder of MgB$_{2}$ (Alfa Aesar)\cite{memo4} was used to make
a pellet. High-pressure heat treatment was performed with a 12-mm cubic
multi-anvil press.\cite{jung7,jung8} The pellet was put into a Au capsule
in a high-pressure cell. A {\it D}-type thermocouple was inserted near the
Au capsule to monitor the temperature. It took about 2 hours to pressurize
the cell to 3 GPa. After the pressurization, the heating power was increased
linearly and then maintained constant for 2 hours. The sample was
sintered at a temperature of $850\sim 950^{\circ}$C and then quenched
to room temperature. The sample weighed about 130 mg and was about 
4.5 mm in diameter and 3.3 mm in height.
The magnetization curves were measured by using a SQUID magnetometer
(Quantum Design, MPMS-{\it XL}).

\section{Results and discussion}
\begin{figure}[b]
\includegraphics[width=7cm,height=5cm]{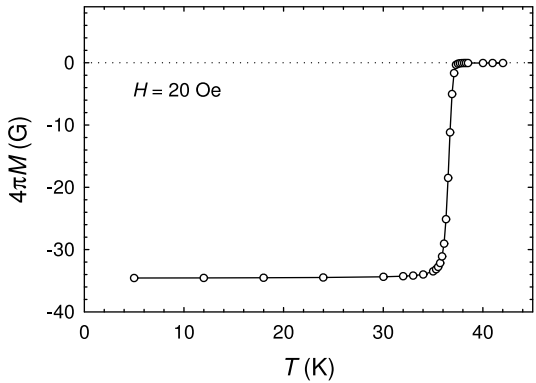}\\
\caption{Zero-field-cooled magnetization, $M(T)$, of MgB$_2$ for $H=20$ Oe.
         This curve reveals the superconducting transition temperature $T_c$
         and the transition width $\Delta T_c$ to be about 37 K and 1 K,
         respectively.}
\label{fig1}
\end{figure}

Figure\ \ref{fig1} shows the temperature dependence of the zero-field-cooled
magnetization measured at $H=20$ Oe. From this figure, we found that
the superconducting transition temperature $T_c$ and the transition
width $\Delta T_c$ were about 37 K and 1 K, respectively.\cite{jung8}

\begin{figure}[tb]
\includegraphics[width=7cm,height=5cm]{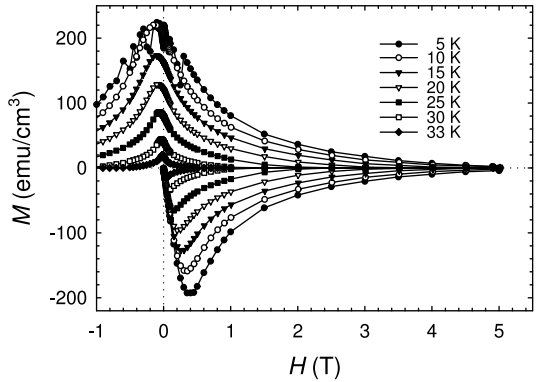}\\
\caption{Magnetization curves, $M(H)$, measured in the region
         5 K $\leq T\leq$ 33 K and $-$5 T $\leq H \leq$ 5 T.}
\label{fig2}
\end{figure}

Figure\ \ref{fig2} shows the magnetization curves, $M(H)$, of MgB$_2$, which were
measured in the temperature region of 5 K $\leq T \leq$ 33 K.\cite{memo5}
One notable feature is the symmetry in the increasing and the decreasing field
branches, {\em i.e.}, $M(H^+)=-M(H^-)$, where $M(H^+)$ and $M(H^-)$ are the
magnetizations in the increasing and the decreasing field branches, respectively.
Such a feature can be commonly found in previous 
reports.\cite{finnemore7,larbalestier7,takano7}
This means that the contribution of the equilibrium magnetization and 
the surface pinning is negligible compared to that of the bulk pinning.
The irreversible magnetization can be described by various critical-state models.
The Bean model\cite{bean1} has been used to calculate the critical current density
of superconducting materials. The model assumes that the slope $dh(r)/dr$
is constant and field independent, where $h(r)$ denotes the local magnetic
induction inside a sample. Thus, the critical current density (or irreversible
magnetization) should  also be field independent, which is contrary to most
experimental results.

Other critical-state models, such as the exponential and the Watson
models,\cite{feitz7,watson7} which take into account the field dependence of
the critical current density, can be used to describe the irreversible
magnetization properly.
In the frame of the Watson model, the critical current density, $j_c(h(r))$, is
given by
\begin{equation}
j_c(h(r))=j_0(1+|h(r)|/h_0),
\end{equation}
where $j_0$ and $h_0$ are adjustable parameters which depend on the material.
The exponential model proposes that the critical current density
has the form
\begin{equation}
j_c(h(r))=j_0\exp(-|h(r)|/h_0),
\end{equation}
where $j_0$ and $h_0$ are again adjustable parameters as in the Watson model.
According to Ampere's law, the field gradient inside a sample is given by
\begin{equation}
\frac{dh(r)}{dr}=-{\rm sgn}(j)\frac{4\pi}{c} j_c(h(r)),
\end{equation}
where sgn($x$) is the sign function and $c$ is the speed of light. 
In cylindrical coordinates,
we obtain an average magnetic induction $\langle h \rangle$ of a sample with a radius $a$
\begin{equation}
\langle h \rangle=B=H+4\pi M=\frac{1}{\pi a^2}\int_0^a\int_0^{2\pi} h(r) d\theta dr.
\label{avginduction}
\end{equation}
If the surface barrier effect is ignored, the boundary condition for $h(r)$
is $h(r=a)=H$, where $H$ is the external magnetic field.

\begin{figure}[tb]
\includegraphics[width=7.8cm,height=9cm]{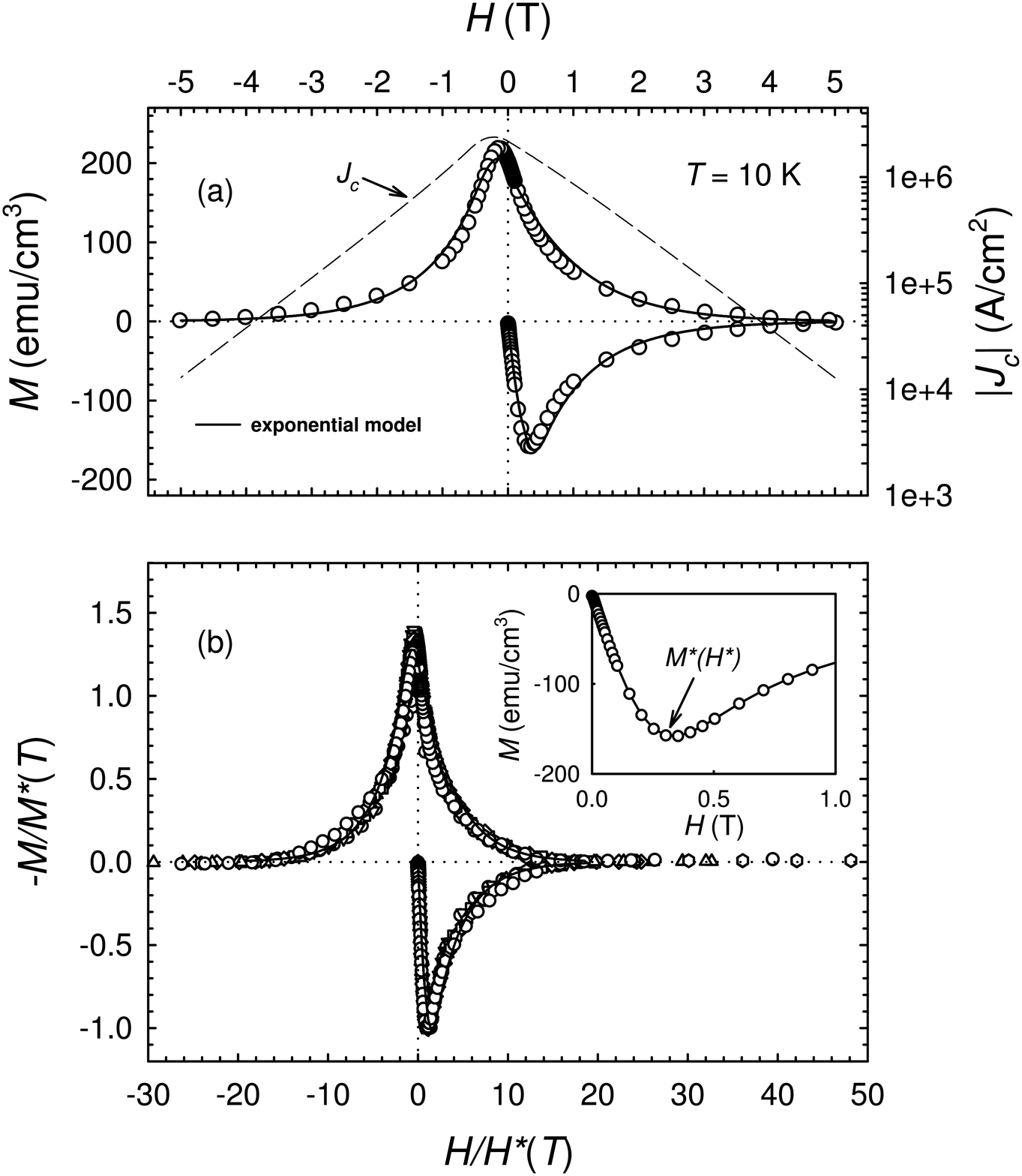}\\
\caption{(a) Magnetization curve, $M(H)$, at $T=10$ K. The solid line represents
        the theoretical curve for the exponential critical-state model. The line
        denotes the absolute value of the average critical current density 
        $J_c(H)=\langle j_c(h(r)) \rangle$ calculated from the decreasing-field 
        branch of the theoretical magnetization curve at $T=10$ K.
        (b) Scaling of the magnetization curves, $M(H)$,in the temperature region
        5 K $\leq T\leq$ 33 K.  The inset illustrates the definitions of $M^\ast$ 
        and $H^\ast$ (see text).}
\label{fig3}
\end{figure}

Figure\ \ref{fig3}(a) shows our attempt
to fit $M(H)$ at $T = 10$ K by using Eq.\ (\ref{avginduction}) with the exponential
critical-state model with $j_0a = 697$ emu/cm$^3$ and $h_0=0.93$ T.
For the theoretical description of the $M(H)$, we can choose an arbitrary number for 
a sample size, $a$, within the constraint that the multiplier $j_0a$ is a constant.
As one can see, the data are well described by the critical-state model
without considering the contribution of the reversible magnetization
and the surface barrier effect.
As stated before, this implies that, in the mixed state in the MgB$_2$ superconductor,
the magnetization mainly comes from the contribution of the bulk pinning effect.
A fit using the Watson model was also attempted, but was poor for
all adjustments of the parameters. The dashed line of Fig.\ \ref{fig3}(a) represents
the average critical current density $J_c(H)=\langle j_c(h(r)) \rangle$ calculated from the 
decreasing-field branch of the theoretical magnetization curve assuming 
the grain size $a=25$ $\mu$m.

We note that the shapes of the $M(H)$ curves shown in Fig.\ \ref{fig2}
are remarkably similar to each other. This suggests that the vortex pinning mechanism 
in MgB$_2$ does not change even as the temperature is varied up to near $T_c$.
More concrete evidence for this
can be found from the scaling analysis of the $M(H)$ curves. For the scaling of the
$M(H)$ curves, we define
two phenomenological parameters as shown in the inset of Fig.\ \ref{fig3}(b).
$H^\ast(T)$ means the field where the magnetization in the increasing field branch
reaches it's maximum value $M^\ast(T)$. We divided the $M$ and the $H$ in each curve of
Fig.\ \ref{fig2} by $-M^\ast(T)$ and $H^\ast(T)$, respectively. The result is shown in
Fig.\ \ref{fig3}(b). Without any exception, all the curves collapse on a single
universal curve. The solid line in the figure denotes the exponential critical-state
model. This result is consistent with the scaling of the pinning force, 
$F_p(H)\propto HJ_c$, in a temperature range of $T\geq 0.5T_c$ reported by 
Larbalestier {\em et al}.\cite{larbalestier7}
In the case of high-$T_c$ cuprates, such scaling behavior of the $M(H)$ curves
is established in a limited low-temperature region. This implies
that the fundamental mechanism determining the magnetic hysteresis
at low temperatures changes as the temperature is increased toward $T_c$.
For Bi$_2$Sr$_2$CaCu$_2$O$_8$ (Bi-2212),
while the bulk pinning is dominant at low temperatures, the contribution of the surface
or geometrical barrier effects to the magnetization becomes more important as the
temperature is increased.\cite{dewhurst7} Thus, universal scaling of $M(H)$ is
not seen in Bi-2212.

\begin{figure}[tb]
\includegraphics[width=7.8cm,height=5cm]{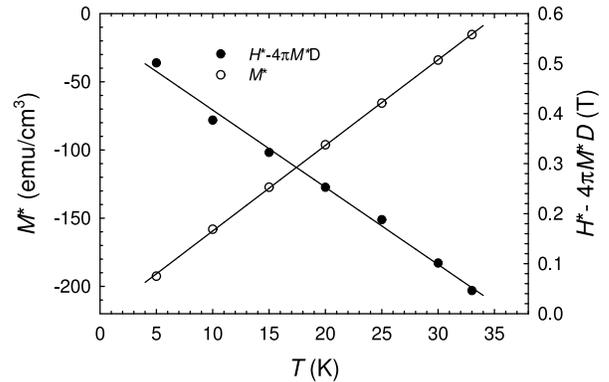}\\
\caption{Temperature dependence of  phenomenological parameters
         $M^\ast$ and $H^\ast-4\pi M^{\ast}D$, 
         where $D$ is the demagnetization factor of the sample. Solid lines 
         represent linear least-squares fits of the data.}
\label{fig4}
\end{figure}

Figure\ \ref{fig4} shows the temperature dependence of the scaling parameters 
$H^\ast(T)$ and $M^\ast(T)$ used in the above analysis. The right axis for $H^\ast(T)$ 
in the figure was corrected by the demagnetization factor $D=0.42$. 
The value of $D$ was obtained from the low-field susceptibility curve 
in Fig.\ \ref{fig1} assuming 100 \% magnetic screening. It was obvious that the 
$M^\ast$ and $H^\ast(T)-4\pi M^\ast(T)D$ scaled lineally with temperature 
as indicated by solid lines. The linearity in $H^\ast(T)-4\pi M^\ast(T)D$ 
requires a linear temperature dependence of the irreversible field $H_{\rm irr}$ 
where the magnetic hysteresis disappears. This is because the normalized hysteresis 
curves $M(H)$ for $T \leq 0.9T_c$ collapsed into a single universal curve 
as we showed. In fact, a nearly linear $H_{\rm irr}(T)$ was shown from 
the Kramer analysis of magnetization.\cite{larbalestier7} This feature differentiates 
MgB$_2$ from other cuprate high-$T_c$ materials with $H_{\rm irr} \sim (T_c-T)^{1.5}$.

\section{summary}
In summary, we measured the magnetization curves $M(H)$ for the newly discovered
metallic MgB$_2$
superconductor, which has a $T_c\simeq37$ K, in the region 5 K $\leq T\leq$ 33 K 
and $-$5 T $\leq H \leq$ 5 T.
The magnetic hysteresis in our experimental region was well described by the
exponential critical-state model, without considering 
the reversible magnetization and the surface barrier effect.
Also, we found that all the magnetization curves collapsed onto a single universal curve
when the field and the magnetization were normalized by the characteristic
scaling factors $H^\ast(T)$ and $M^\ast(T)$, respectively.
These results lead us to the conclusion that the irreversible magnetization of MgB$_2$
is dominated by bulk pinning and that the pinning mechanism does not change even when
the temperature is varied up to $T/T_c\sim 0.9$.

\acknowledgements
This work was supported by the Ministry of Science and Technology of Korea
through the Creative Research Initiative Program.


\end{multicols}
\end{document}